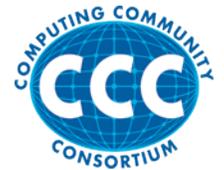

# From Data to Knowledge to Action: Enabling the Smart Grid

**Randal E. Bryant**
Carnegie Mellon University

**Chase Hensel**
Computing Research Association

**Randy H. Katz**
University of California-Berkeley

**Erwin P. Gianchandani**
Computing Research Association

Version 5:  September 21, 2010[1]

Our nation's infrastructure for generating, transmitting, and distributing electricity – "The Grid" – is a relic based in many respects on century-old technology.  It consists of expensive, centralized generation via large plants, and a massive transmission and distribution system.  It strives to deliver high-quality power to all subscribers simultaneously – no matter what their demand – and must therefore be sized to the peak aggregate demand at each distribution point. Power is transmitted via high voltage lines over long distances, with associated inefficiencies, power losses, and right-of-way costs; and local distribution, via step-down transformers, is expensive in cost and efficiency, and is a single point of failure for an entire neighborhood.  Ultimately, the system demands end-to-end synchronization, and it lacks a mechanism for storing ("buffering") energy, thus complicating sharing among grids or independent operation during an "upstream" outage.

Recent blackouts demonstrate the existing grid's problems – failures are rare but spectacular. Average demand per consumer is a small fraction of the peak – a 25 kWhr/day home draws on average less than five percent of its 100-amp service.  Consumption correlations, e.g., air conditioners on a hot day, drive demand beyond estimated aggregates, which can result in huge spikes in supply cost and may trigger blackouts.  Moreover, the structure cannot accommodate the highly variable nature of renewable energy sources such as solar (generating power only during the day) and wind (generating power only when the wind is strong enough).  Meanwhile, consumers are provided little information about their energy usage (just a monthly total) and even fewer opportunities or incentives to adapt their usage to better align their demands to the capabilities of the utility companies.

Many people are pinning their hopes on the "smart grid" – i.e., a more distributed, adaptive, and market-based infrastructure for the generation, distribution, and consumption of electrical energy.  This new approach is designed to yield greater efficiency and resilience, while reducing environmental impact, compared to the existing electricity distribution system.[2]  Already, the U.S. government is investing billions of dollars in deploying aspects of smart grid technology, primarily "smart" meters and associated communications technology. These meters can display real-time prices to consumers, to motivate them to reduce their consumption at times of high demand. But trepidation about trusting such critical infrastructure to technology that is not yet proven may limit just how "smart" the smart grid can become in the near-term. In addition, little attention is being paid to more radical approaches, ones that would involve fundamentally new, more decentralized structures for the grid.

---

[1] Contact:  Erwin Gianchandani, Director, Computing Community Consortium (202-266-2936; erwin@cra.org).  For the most recent version of this essay, as well as related essays, visit http://www.cra.org/ccc/initiatives.

[2] It is important to note that there is a distinction between saving peak energy (which can be done with load shifting using storage and delayed water heating, for example) and energy savings.  Charging and discharging storage, for example, may help reduce peak but actually increase energy consumption (and $CO_2$ emissions).  The value of a "smart grid" is to help save emissions and energy by ensuring maximum use of low emissions generation (e.g., renewable, nuclear, etc.).

Initial plans for the smart grid suggest it will make extensive use of existing information technology. In particular, recent advances in data analytics – i.e., data mining, machine learning, etc. – have the potential to greatly enhance the smart grid and, ultimately, amplify its impact, by helping us make sense of an increasing wealth of data about how we use energy and the kinds of demands that we are placing upon the current energy grid. In other words, data analytics approaches are facilitating a *data* → *knowledge* → *action* paradigm in the energy space; they are enabling us to generate knowledge from data and make informed, real-time decisions about specific actions, thereby yielding a truly intelligent smart grid.

Here we describe what the electricity grid could look like in 10 years, and specifically how Federal investment in data analytics approaches are critical to realizing this vision.

**The Smart Grid in 2020**

A critical step in enhancing our nation's electricity infrastructure involves smartly extending it into homes, offices, and factories. Consider the following vision for how electricity may be served to a home in 2020:

> *The Jones family, of Phoenix, Arizona, lives in a house with state-of-the art sensing and control capabilities. Their home management computer has collected data on the habits and preferences of the family in order to create a model of their electricity use. The management computer uses this model to inform its decisions and interactions with the electric utility's computer system. On August 15, 2020, the following interchange takes place between these two systems:*
>
> *The utility company sees that the day's temperature will exceed 110°F. It needs to reduce the peak demand later in the day. It contacts the Jones's home management computer (JC) and requests that it keep the load below 8KW between 4pm and 9pm. JC is aware that the family will be out until 8pm, and it knows (a) it can safely let the household temperature rise to 90°F without harming the family cat or tropical fish, and (b) it takes 30 minutes to bring the temperature back down to the Jones family 's preferred reading of 75°F. Consequently, JC offers to reduce usage to 4KW from 4pm to 7:30pm, then to 12KW from 7:30pm to 8pm and to 6KW from 8pm to 9pm. JC can maintain the latter limit because it anticipates that the Joneses will be returning with their electric car still partially charged; JC will be able to tap into the energy remaining in the car's batteries. It bundles this offer with a requirement that the utility company supply enough energy to fully recharge the car by 7am the next day, and it offers to supply energy to the grid from the household solar panels for the rest of the day. The two systems negotiate an appropriate price for the total energy package.*

While the ideas of using pricing incentives and household automation have been proposed as part of the system for achieving more balanced loads and better efficiencies in the smart grid, the above scenario includes aspects that go much further than other existing plans for smart grid technology:

- The home management computer serves as the "energy czar" for the home. It uses sensors to "learn" the usage patterns and characteristics of the household appliances and lighting (e.g., power draw, ramp-up time), as well as the occupants' habits, needs, and preferences beyond default settings. It can selectively control devices, such as the household thermostat. It can diagnose anomalies (e.g., defective fluorescent lighting ballasts or refrigerator doors left ajar) and report these to the homeowners. It has access to the homeowners' calendar programs and is able to determine when they will be away. Unlike some smart grid proposals that require humans to continually monitor and adjust their energy usage, this example illustrates how the home management computer could function much like a good butler — it would automatically learn the occupants' preferences and make the right choices in the background.
- The home management computer negotiates with the utility company, engaging in complex transactions involving bundled combinations of current and future quotas and prices, as both a consumer and a provider of electricity.
- The utility company's system must perform these negotiations with hundreds of thousands of households, as well as with hundreds of other entities such as power generators, transmission line operators, other utility companies, etc. The system has at its disposal data about weather and seasonal patterns, but it must also consider possible statistical variations and unexpected demands and outages.
- Although we described the above scenario as involving a single negotiation per day, it is likely that utility companies and home computer systems will be engaging in these exchanges more frequently (e.g., hourly).

**Data analytics as a driver**

Much of the technology underlying such a system requires advances in the area of data analytics, including:

- *Machine learning/data mining* can readily detect usage patterns and preferences automatically. For example, researchers at Georgia Tech and the University of Washington have shown that instrumenting a home with just three sensors — for electric power, water, and gas — makes it possible to determine the resource usage of individual appliances, lights controlled by individual switches, and the HVAC and plumbing systems — simply by analyzing patterns in the data. The sensor and control system can apply machine learning to continually improve energy efficiency, reliability, and comfort by monitoring operations and algorithmically tuning parameters and behaviors, largely eliminating the need for users to manually set configuration parameters. Realizing this potential requires that sensors and the control system work reliably in all environments and at sufficiently low cost for a consumer marketplace.
- *Agent-based (auction) systems* can negotiate complex contracts on a massive scale. For example, Google performs automated auctions millions of times per day to place paid advertisements within its search result pages. But it's one thing to sell space on Web pages;

it's a much more serious proposition to operate a market-based system at this scale that controls critical infrastructure.
- *Advanced optimization* can guide the adoption of renewable energy sources, such as wind and solar, based on projected macro-scale demand, grid capacity with anticipated upgrades, and consideration of the inherent intermittency of renewable power sources. For instance, the optimal location in terms of wind-energy density may not be as desirable as a slightly suboptimal location where projections indicate maximal need; two smaller wind farms on opposite sides of a geographical barrier (e.g., a mountain range) may prove most efficient due to offsetting intermittency, reasonable grid access, and consistency with planned grid upgrades.

**Some challenges**

Fully realizing a future in which millions, perhaps billions, of computerized agents manage our energy production, distribution and consumption requires many capabilities well beyond those of current data analytics tools and consequently of any system relying upon them. Indeed, moving to the smart grid of the future presents many fundamental challenges that we have yet to address:

- *The systems must adapt to unexpected events*. What if the Jones' return home early, or their car returns with less charge in its batteries than was anticipated? Many other events can undo the careful planning made by the utility company: new usage patterns, unexpected weather conditions, failures of components or subsystems, etc. The systems must operate with sufficient capacity margins to avoid failures. The agents must be able to dynamically renegotiate contracts, with appropriate pricing mechanisms to avoid abuse.
- *The system components must be able to cooperate with one another*. For example, shouldn't the Jones' home computer be able to remotely query the Jones' car during the course of the day to assess the precise level of charge anticipated upon the Jones' return, and could the Jones' home computer in turn use this information to renegotiate contracts in "real-time"? This high degree of connectivity and coordination could make the system more reliable; however, if poorly designed, the system could also be more vulnerable to cascading failures leading to large-scale blackouts.
- *The system must guarantee sufficient privacy*. The Jones family might not want the utility company (or a malicious eavesdropper) to know things like when the house is vacant or when the teenage daughter is home alone. Unlike scenarios where utility companies are provided direct control over household appliances, we envision that the home management computer will serve as an "information firewall" to the outside world. It will act on behalf of the homeowners while restricting the flow of information to the outside world. It may even choose to obfuscate externally visible usage patterns, e.g., by having some form of energy storage within the house that can be charged or utilized at different times of day. (As with many other real-world systems, there may be a benefit-cost tradeoff between privacy and efficiency.) Ultimately, these technologies will require computer scientists interfacing with policymakers directly.
- *The system must be resilient to abuse or attack*. Experience with the California energy market in 2000 demonstrated the possibility for companies to "game the system," creating havoc while reaping huge monetary benefits by exploiting flaws in the computerized marketplace. Given the rise in the amount and sophistication of Internet-/cybercrime, there are justifiable fears that malicious agents will target any network-based smart grid both for monetary gain and to disrupt the U.S. economy. An effectively designed, agent-based system can potentially be less vulnerable to manipulation or attack than a centralized, monolithic one, but a bad design could yield just the opposite effect.
- *The system must learn and improve over time.* As new electrical devices become available, how should they be incorporated into the power optimization equation most efficiently? As usage patterns evolve with households becoming more energy conscious, or as families evolve (e.g., children are born, or children go off to college) and so do their energy consumption patterns, the underlying machine learning must track and adapt, both to long-term lifestyle changes and to transient ones (e.g., a family goes on a two-week vacation, or workmen remodel a home and their power tools draw substantial electricity for a short time).

**Other smart grid areas**

Although the example and discussion presented above focus mainly on residential electricity, many of the ideas embedded therein can be extended to all forms of energy generation, distribution and usage, including transportation. For example, critical savings of energy come from other kinds of sophisticated controls in buildings and vehicles. In buildings, these controls include the obvious (e.g., turning off lights when no one is in the room, adjusting artificial light levels based on the amount of ambient light, installing electrochromic windows, adjusting ventilation levels to meet the actual needs of each space, etc.) to the more subtle: anticipating when someone is about to enter a space; or cooling a building core in the early morning when outdoor temperatures are low and chilling equipment is most efficient. It is critical to address these subtleties by building technologies that are able to learn about a given system's (and/or occupant's) behavior and to update this analysis as situations change. Similar strategies apply for vehicles. In both cases, learning is essential to optimize design.

The systems also need to be alert for anomalies that might be caused by failing sensors or equipment. For example, studies in commercial buildings have demonstrated how simply reattaching or fixing failed sensors, fixing broken motors, unclogging air intake flaps, retuning the control software, etc., can net up to 30 percent in energy savings. New techniques for sensing such failures – either at the source of the failure or through analysis of energy consumption patterns enabling visualization and detection of faults – are therefore essential. For example, researchers are already installing sensor networks into office spaces in order to monitor energy usage – and analysis of the resultant data is highlighting problems, such as both the heating and cooling system being on at the same time, offices that keep lights on even when workers are absent, etc. In a few test cases in the Bay Area of California, buildings have significantly diminished the amount of electricity they waste on a daily basis based on knowledge learned – and actions taken – as a result of these data analytics approaches.

Irrespective of the efficiency gains caused by the use of regenerative methods converting kinetic energy back into electricity (e.g., regenerative breaks and shocks), the charge/discharge pattern of urban electric vehicles has lots of peaks and valleys. This type of charge/discharge profile is bad for batteries, as it results in an artificial shortening of batteries' lifespan and efficiency. Ideally, a battery should be slowly, and smoothly, charged and discharged. Some researchers are installing supercapacitors – electric storage devices well suited for urban chaos – as buffers between the battery, the motor, and the regenerative system. Early results show that – given the proper signal analysis and control algorithm – an affordable supercapacitor can improve urban car efficiency by over 30 percent and increase battery longevity significantly.

Finally, the behavioral aspects of providing consumers with better information to make informed choices (including ways of benchmarking their energy usage versus that of others) constitute separable and potentially powerful tools. Consider healthcare, where we are witnessing transformations in how care is being delivered through the use of social media like *Patients Like Me* that allow persons afflicted with chronic disease to connect with others like them, compare vital signs and test results, and assess which care solutions might be best. More and more, *patients are recommending to their doctors* specific treatments, in stark contrast to the way medicine has been practiced for centuries. Likewise, new technologies are being developed to capitalize on ubiquitous computing capabilities, with researchers analyzing large medical and behavioral data sets to generate mobile phone-based applications that facilitate behavior change – toward healthier living. Similar strategies could be adopted in the energy space. For example, a mobile phone could sense when its user leaves his or her office at the end of a busy workday – based on prior patterns, GPS systems, etc. – and remind the individual to turn off the lights.

**Leadership and research funding**

Our country lacks any central organization, whether in the private sector or within the government, that has either the authority or the motivation to make major changes in our electric grid. Indeed, the utility companies operate as regulated monopolies with limited incentive to innovate. For example, much of the Federal funding for utility companies from the American Recovery and Reinvestment Act of 2009 has been spent on buying smart meters, not on supporting new research and development efforts for identifying tomorrow's technological innovations today. Other countries, including Denmark, Spain, and Brazil, are more advanced in their use of renewable energy sources. Numerous industry-driven smart grid initiatives have been undertaken, yet these have been fairly incremental in nature, and none has yielded a comprehensive system architecture for the future grid. Nevertheless, as the home of much of the information technology innovations of the last half-century, the U.S. is well positioned to benefit from information technology-enabled energy efficiency. Undoubtedly, a focus on this effort plays to America's technological strengths. **Thus, the Federal government should take the initiative in this area by (a) funding research into and development of proof-of-concept prototypes, (b) organizing testbeds and partnerships between the information technology and energy technology industries, and (c) developing new research communities at the intersection of data analytics and energy systems.**

Current research funding for smart grid technology is highly fragmented across the Federal government, to the detriment of the collaborative and interdisciplinary approach that is necessary in order to appropriately address the many challenges that must be overcome in this area. For example, 84 percent of all Federal funding for basic research in computer science comes from the National Science Foundation's Directorate for Computer and Information Science and Engineering (CISE), but NSF on its own cannot be expected to deal with the broad set of issues that must be addressed to develop and deploy smart grid technology. NSF/CISE does have plans to fund research in sustainable energy through a NSF-wide Science, Engineering, and Education for Sustainability (SEES) initiative included in the President's FY 11 budget request to Congress (see www.nsf.gov/sees).  SEES promises to create interdisciplinary collaborations among researchers across the disciplines supported by the NSF to generate "decision capabilities and technologies aimed at mitigating and adapting to environmental change that threatens sustainability."  However, several factors will limit the impact this program can have on smart grid technology.  First, it will span a broad range of topics, such as environmental protection and climate modeling, and hence the portion that addresses energy issues will be just a fraction of the total.  Second, the U.S. Department of Energy serves as the locus of activity for energy-related technology and policy; consequently, with NSF as the funding source for SEES, it is not likely that the full complement of engineers, scientists, and policy researchers, across academia, industry, and government, will be at the table to work together on addressing the complexity of creating and implementing smart grid technology.

The natural home for smart grid research, the U.S. DoE, has to date provided only minimal funding to the core data analytics components.  For example, the initial solicitation by the newly-established Advanced Research Projects Agency-Energy (ARPA-E) in 2009 led to 37 funded projects, but only one – a Stanford project on providing consumers with better information about their energy usage – has a significant data analytics component.  More recent solicitations from ARPA-E appear to narrow the range of topics and even further reduce support for data analytics research.  The DoE Office of Science has funded computing research in the past, but this work has been limited to high performance computing, simulation, and modeling.  While these are certainly important and relevant research areas, they fail to touch the data analytics issues underlying smart grid technology discussed in this report.

**We feel it is imperative for the U.S. DoE, and perhaps in conjunction with NSF/CISE, to provide avenues of support for highly collaborative, multi-disciplinary teams comprising computer scientists that seek to address the challenges of our energy infrastructure.**  At the very least, we have provided justification here for ARPA-E to include strong data analytics components in its current and future solicitations.  In addition, other potential partners in this space include:

- **DoE's Office of Electricity**, which has the lead for smart grid activities and is supporting a variety of projects in this space.

- **DoE's Office of Energy Efficiency and Renewable Energy (EERE)**, which is working on a number of simulations trying to understand linkages between different patterns of demand, transmission infrastructure, demand side management, storage (in different locations and at different scales), large penetrations of wind and solar power, etc.  For example, currently, the National Renewable Energy Laboratory (NREL) is building an entirely new test lab called the Energy Systems Integration Facility (ESIF) to combine models with actual high power equipment tests of inverters, storage devices, etc.

- **The National Institute of Standards and Technology (NIST)**, which has taken the lead in defining smart building integration standards.  A key hurdle to the vision described above is interoperability of equipment in buildings and other systems.  For example, a new water heater must be "plug and play" with the other systems already installed in a home. Consequently, any basic research must consider the standards NIST is establishing.

The efforts of all of these Federal offices/agencies would strongly benefit from basic advances in data analytics through the efforts of computing researchers.

**The road ahead**

The director of ARPA-E recently wrote, "The nation that successfully grows its economy with more efficient energy use, a clean domestic energy supply, and a smart energy infrastructure will lead the global economy of the 21$^{st}$ century. In many cases, [the U.S.] is lagging behind. We as a nation need to change course with fierce urgency." Achieving a truly smart energy infrastructure – for energy generation, distribution, and consumption – inherently requires basic and advanced computing research, as outlined above. Today, computer scientists are well-equipped to collaborate with other scientists and engineers, enabling the current concepts for the smart grid to be realized and then taking the vision to entirely new levels, yielding fundamental improvements in efficiency, reliability, and security, all the while reducing environmental impact. We can no longer afford to wait for this work to get underway.